\RequirePackage{fix-cm}
\documentclass[twocolumn]{svjour3}

\usepackage{mathptmx, amsmath, amssymb, amsfonts, hyperref, graphicx}

\usepackage[ruled,vlined]{algorithm2e} 

\SetKwBlock{Begin}{begin}{end} 
\SetKwInput{KwIn}{input}

\newcommand{\f}{\mathbf f}
\newcommand{\h}{\mathbf h}
\newcommand{\x}{\mathbf x}
\newcommand{\M}{\mathbf M}
\newcommand{\X}{\mathfrak X}
\newcommand{\eff}{\mathrm{eff}}
\newcommand{\RR}{\mathbb R}

\journalname{}

\begin{document}

\title{Optimal Design of Multifactor Experiments via Grid Exploration}

\author{Radoslav Harman \and
        Lenka Filov\'{a} \and
        Samuel Rosa
}

\institute{R. Harman, L. Filov\'{a}, S. Rosa \at
           Department of Applied Mathematics and Statistics\\
           Faculty of Mathematics, Physics and Informatics\\
           Comenius University in Bratislava, Slovakia\\
           \email{harman@fmph.uniba.sk}     
}

\date{Received: date / Accepted: date}

\maketitle

\begin{abstract}
For computing efficient approximate designs of multifactor experiments, we propose a simple algorithm based on adaptive exploration of the grid of all combinations of factor levels. We demonstrate that the algorithm significantly outperforms several state-of-the-art competitors for problems with discrete, continuous, as well as mixed factors. Importantly, we provide a free \texttt{R} code that permits direct verification of the numerical results and allows the researchers to easily compute optimal or nearly-optimal experimental designs for their own statistical models.
\end{abstract}

\section{Introduction}\label{sec:Introduction}

The usual aim of the so-called ``optimal'' design of experiments is to perform the experimental trials in a way that enables efficient estimation of unknown parameters of an underlying statistical model (e.g., \cite{Fedorov}, \cite{Pazman}, \cite{Puk}, \cite{Atkinson}, \cite{GJ}, \cite{PronzatoPazman}). The literature provides the analytic forms of optimal designs for many particular situations; for the practical problem at hand, however, the analytic results are often unavailable. In that case, it is usually possible to compute an optimal or nearly-optimal design numerically (e.g., Chapter 4 in \cite{Fedorov}, Chapter 5 in \cite{Pazman}, Chapter 12 in \cite{Atkinson}, and Chapter 9 in \cite{PronzatoPazman}).

In this paper, we propose a simple algorithm for solving one of the most common problems within this area: computing efficient approximate designs for experiments with uncorrelated observations and several independent factors. The proposed algorithm employs a specific strategy to adaptively explore the grid of factor level combinations, without the need to enumerate all of the elements of the grid. The key idea of this algorithm is to form exploration sets composed of star-like subsets and other strategically selected points; therefore, we will refer to this algorithm as the ``galaxy'' exploration method (GEX).

If the set of all combinations of factor levels is finite and not too large, it is possible to use many available efficient and provably convergent algorithms to compute an optimal design (e.g., \cite{Fedorov}, \cite{Atwood73}, \cite{STT}, 
\cite{Bohning}, \cite{VBW}, \cite{SIMD}, \cite{Yu}, \cite{Sagnol}, \cite{YBT}, \cite{REX}). However, in the case of multiple factors, each with many levels, the number of factor level combinations is typically much larger than the limit of the applicability of these methods. 

The main advantage of GEX is that it can solve problems with an extensive number of combinations of factor levels, e.g., $10^{15}$ ($5$ factors, each with $1000$ levels), and obtain an optimal design in several seconds. Note that factors with large, yet finite, numbers of levels often correspond to practical requirements on the factor resolution. Even in the theoretical case of fully continuous factors, GEX can be applied; the factors can simply be densely discretized (to, say, $4$ decimal places). We will show that this straightforward approach usually significantly outperforms the intricate state-of-the-art methods that are allowed to choose the factor levels anywhere in given continuous intervals (e.g., \cite{PZ}, \cite{Duarte}, \cite{Garcia-Rodenas}, \cite{Lukemire}, \cite{Xu}). As a by-product, we demonstrate that it is rarely necessary to discretize each factor to an enormous number of levels to admit an approximate design that is almost perfect compared to the optimum achievable with fully continuous factors. 
\bigskip

Let $k$ be the number of experimental factors. Suppose that for any trial the factors can be independently chosen within pre-defined finite sets $\X_1,\ldots,\X_k \subset \RR$ of permissible levels, without mutual constraints. The design space $\X$ is then the set of all combinations of the factor levels, formally $\X = \X_1 \times \cdots \times \X_k$. Geometrically, $\X$ is a finite grid of ``design points'' $\x=(x_1,\ldots,x_k)^T$. 

An approximate experimental design is any probability measure $\xi$ on $\X$. For $\x \in \X$, the interpretation of $\xi(\x)$ is the proportion of trials to be performed in $\x$. The value $\xi(\x)$ is often referred to as the ``weight'' of the design point $\x$. The set of all design points $\x$ with non-zero weights is called the support of $\xi$ and denoted $\mathrm{supp}(\xi)$. 

In practical experiments, an approximate design $\xi$ must be converted to an ``exact'' design of size $N$ determined by available experimental resources.\footnote{This holds if each trial consumes a constant amount of the resources, independent of other trials. The situation is fundamentally more complex under general restrictions on permissible experimental designs, see, e.g., \cite{HBF}, \cite{SagnolHarman}, \cite{AQUA}.} The exact design assigns non-negative integer numbers $n_1,\ldots,n_s$ of trials to properly selected points $\x_1,\ldots,\x_s$, such that $\sum_{i=1}^s n_i=N$. A standard strategy is to select the points $\x_1,\ldots,\x_s$ that form the support of $\xi$, and the integer numbers of trials are then computed by appropriate rounding of the numbers $N\xi(\x_1),\ldots,N\xi(\x_s)$, see, e.g., \cite{PukelsheimRieder}. Alternatively, since the support size $s$ is usually small, it may be feasible to use specialized procedures for optimal exact designs (e.g., Chapter 12 of \cite{Atkinson}) on $\mathrm{supp}(\xi)$ to compute the numbers $n_1,\ldots,n_s$. Note that if we adopt this alternative approach then the required number $N$ of trials can be both larger and smaller than the identified support size $s$.

Approximate designs lead to an optimization problem that is generally much simpler than a direct computation of efficient exact designs of a given size $N$. Moreover, the approximate optimal design, once computed, can be used for the construction of an exact designs of any size; for a moderate and large size of the experiment, these exact designs tend to be very efficient. In addition, the optimal approximate design can be utilized to provide informative lower bounds on the quality of candidate exact designs. In the rest of this paper, we will use the term ``design'' to specifically denote an approximate experimental design.
\bigskip

The quality of a design $\xi$ is usually expressed as a function of the ``normalized'' information matrix
\begin{equation}\label{eq:infmat}
  \M(\xi)=\sum_{\x \in \X} \xi(\x) \f(\x)\f^T(\x),
\end{equation}
where $\f: \X \to \RR^m$, $m \geq 2$, is known.\footnote{Note that in the applications of optimal designs the dimension $m$ is usually less than $10$.} To avoid uninteresting pathological situations, we will assume that $\f(\x_1),\ldots,\f(\x_m)$ are linearly independent for some $\x_1,\ldots,\x_m \in \X$. The information matrix of the form \eqref{eq:infmat} is typical of models with independent, real-valued observations with non-zero variance, where the stochastic distribution of each observation $y$ depends on the design point $\x$ chosen for the corresponding trial, as well as on the unknown $m$-dimensional vector $\theta$ of model parameters. 

For linear regression, i.e., $\mathrm{E}(y(\x))=\h^T(\x)\theta$, $\mathrm{Var}(y(\x))=\sigma^2/w(\x)$, where $\h:\X \to \RR^m$, $w: \X \to (0,\infty)$ are known and $\sigma^2$ is possibly unknown, we have
\begin{equation*}
\f(\x)=\sqrt{w(\x)}\h(\x).
\end{equation*}
However, if $\mathrm{E}(y(\x))=\eta(\x,\theta)$, where $\eta: \X \times \RR^m \to \RR$ is non-linear in $\theta$, the situation is generally much more difficult. In this paper we will adopt the typical simplification, called the approach of ``local'' optimality which makes use of a ``nominal'' parameter $\theta_0$, assumed to be close to the true value (see, e.g., \cite{Chernoff} or Chapter 17 in \cite{Atkinson} and Chapter 5 in \cite{PronzatoPazman}). If $y(\x)$ is normally distributed with possibly unknown constant variance $\sigma^2$, and $\eta(\x,\cdot)$ is differentiable, the approach of local optimality leads to 
\begin{equation*}
  \f(\x) = \left. \frac{\partial \eta(\x,\theta)}{\partial \theta} \right|_{\theta = \theta_0}.
\end{equation*}
For the generalized linear model (GLM) with the mean value of observations $\eta(\x,\theta)=g^{-1}(\h^T(\x)\theta)$, where $g: \RR \to \RR$ is a strictly monotone and differentiable link function and $\h: \X \to \RR^m$ is known, we have
\begin{equation*}
 \f(\x) = \sqrt{w(\x,\theta_0)} \h(\x)
\end{equation*} 
for some $w:\X \times \RR^m \to (0,\infty)$. In Table \ref{tab:models} we provide the form of $w$ for several common classes of GLMs (for more details on computing optimal designs in GLMs, see, e.g., \cite{GLM} and \cite{AW}).

\begin{table}
\caption{Selected GLM classes and the corresponding functions $w(\x,\theta_0)$. The symbol $\varphi$ denotes the standard normal density and $\phi$ denotes the standard normal cumulative distribution function.}\label{tab:models}
\begin{tabular}{llll}
\hline\noalign{\smallskip}
 GLM class (distribution) & Link $g(\eta)$ & Function $w(\x,\theta_0)$ \\
 \noalign{\smallskip}\hline\noalign{\smallskip}
 Logistic (Bernoulli) & $\log\frac{\eta}{1-\eta}$ & $\frac{e^{\h^T(\x)\theta_0}}{(1+e^{\h^T(\x)\theta_0})^2}$ \\ 
 Probit (Bernoulli) & $\phi^{-1}(\eta)$ & $\frac{\varphi^2(\h^T(\x)\theta_0)}{\phi(\h^T(\x)\theta_0)(1-\phi(\h^T(\x)\theta_0))}$ \\
 Poisson (Poisson) & $\log\eta$ & $e^{\h^T(\x)\theta_0}$\\
 \noalign{\smallskip}\hline
\end{tabular}
\end{table}

The size of an information matrix (and, implicitly, the quality of the corresponding design) is measured by an optimality criterion $\Phi$. For clarity, we will focus on the most common criterion of $D$-optimality, but the method proposed in this paper can be trivially adapted to a large class of criteria. The $D$-optimal design problem is to find a design $\xi^*$ that maximizes
\begin{equation*}
\Phi(\M(\xi)) = {\det}^{1/m}(\M(\xi))
\end{equation*}
over the set $\Xi$ of all designs. The $D$-optimal design minimizes the volume of the confidence ellipsoid for the vector of unknown parameters.\footnote{For non-linear models, this statement is valid asymptotically.} The criterion $\Phi$ is concave, positive homogeneous and Loewner isotonic on the set of all $m \times m$ non-negative definite matrices. Note also that the set $\Xi$ is convex and compact, that is, the problem of $D$-optimal design is convex and always has at least one optimal solution. The information matrix of all $D$-optimal designs is non-singular and unique, albeit for some models, even those that do appear in practice, the $D$-optimal design itself is not unique. 
\bigskip

The $D$-efficiency of a design $\xi$ relative to a design $\zeta$ with $\Phi(\M(\zeta))>0$ is defined as $\eff(\xi|\zeta)=\Phi(\M(\xi))/\Phi(\M(\zeta))$. Efficiency of a design $\xi$ (per se) means the efficiency of $\xi$ relative to the $D$-optimal design $\xi^*$. The efficiency of $\xi$ satisfies (e.g., \cite{Puk}, Section 5.15.)
\begin{equation}\label{eVarFunction}
\eff(\xi|\xi^*) \geq \frac{m}{\max_{\x \in \X} d_\xi(\x)},
\end{equation}
where $d_\xi(\x) = \f^T(\x)\M^{-1}(\xi)\f(\x)$. The function $d_\xi(\cdot)$ is called the ``variance function'' because it is proportional to the variance of the least squares estimator of $\f^T(\cdot)\theta$ in the linear regression model. Equation \eqref{eVarFunction} implies the so-called equivalence theorem for $D$-optimality: A design $\xi$ is $D$-optimal if and only if $\max_{\x \in \X} d_\xi(\x) =m$; otherwise, $\max_{\x \in \X} d_\xi(\x) > m$ (see \cite{Kiefer}). Therefore, provided that the maximum of $d_\xi(\cdot)$ over $\X$ can be reliably determined, it is possible to construct a lower bound on efficiency of $\xi$ or prove its optimality. However, if $\X$ is large and multidimensional, the maximization can be challenging, because $d_\xi(\cdot)$ is typically non-linear and multi-modal, even for linear regression models.  

In the rest of the paper, we will drop the prefix ``$D$-'', e.g., a $D$-optimal design will be called just an ``optimal design''. 
\bigskip

We are not aware of any method that specifically targets optimal design problems on very large discrete grids. For the case of factors that are discrete with a huge number of practically possible levels, the  usual approach is to consider the factors to be fully continuous. Therefore, the methods that operate on continuous spaces are natural competitors for the method proposed in this paper.

If one or more factors are assumed to be continuous, the simplest approach is to discretize the factors and turn the problem into one that can be solved by discrete-space methods. Our experience shows that the loss in efficiency is usually negligible if a continuous factor is discretized to, say, $1000$ levels. Such a discretization can be handled with modern hardware and state-of-the-art discrete-space methods for problems with one or two continuous factors, or problems with one continuous factor and a few binary factors.

There are also other situations with continuous factors where it is possible to directly apply the existing discrete-space methods. For instance, for some models there exist theoretical results limiting the support of an optimal design to a relatively small finite subset. As an example, for the fully quadratic linear model on the cube, the search can be restricted to the set of centres of the faces of all dimensions (see \cite{Heiligers}).  

However, the direct discretization may not be good enough if there are more than two continuous factors or if we require a fine resolution of the levels. In such cases, choosing a method such as the one proposed in this paper can be of significant benefit. 

There are many methods for computing optimal designs if all or some factors are continuous. To give a brief survey, we decided to split them into two categories.

\paragraph{1) The methods that solve the optimal design problem using standard continuous-space algorithms.}{ The problem of optimal design with continuous factors can be directly solved by common non-linear programming methods; e.g., the Nelder-Mead algorithm is used in \cite{CL89}, a quasi-Newton method in \cite{Atkinson} (see Section 13.4), semi-infinite programming in \cite{DuarteWong}, and multiple non-linear programming methods are applied in \cite{Garcia-Rodenas}. Some papers utilize non-linear programming methods in an specific way or for a particular class of models; see for instance \cite{GribikKortanek} and \cite{Papp}.

In addition to the classical nonlinear programming methods, various metaheuristics have been recently proposed to compute optimal designs with continuous factors; see  \cite{WCHW} for the particle swarm optimization, \cite{Xu} for the differential evolution, \cite{Lukemire} for the quantum-behaved particle swarm optimization, and \cite{ZWT} for the competitive swarm optimizer and other papers. A thorough empirical evaluation of the performance of non-linear programming methods and metaheuristics, including the genetic algorithms, is given by \cite{Garcia-Rodenas}.

Typical of the above-mentioned methods is that a fixed-size support of the design is merged with the vector of weights to form a single feasible solution of the optimization problem. Then, the algorithms simultaneously adjust the support and the corresponding weights. A major advantage is that this approach can be used with almost any criterion; it is only required to implement the evaluation of the criterion as a subroutine. A disadvantage is that these methods either require a search for an appropriate size of the support, or they are rendered less efficient by using a relatively large support size provided by the Carath\'{e}odory theorem (e.g., \cite{Pazman}). In addition, the convergence to a good design is not guaranteed and, even in the case of convergence, it can be slow. One reason for the slow speed is that the direct application of non-linear programming and meta-heuristics typically does not exploit the convexity of the problem in the design weights.}

\paragraph{2) The methods that use a discrete-space solver together with an idea of ``adaptive grids''.}{An alternative idea is to use a discrete-space method applied to a finite set, which is sequentially updated in the continuous space to (hopefully) cover the support of the optimal design. We will call this approach ``adaptive support''. The general idea of an adaptive search for the optimal support has been present in the literature from the beginning; for instance, the classical Fedorov-Wynn algorithm is based on iterative improvements of the support of the design, although at each new support the design weights are only adjusted by a simple transformation (\cite{Fedorov}, Chapter 4). The general possibility of using adaptive supports has also been mentioned in \cite{Wu} and \cite{Wu78}. Similarly, in \cite{SJ12}, Section 6, the authors propose ``...a multi-stage grid search that starts with a coarse grid that is made increasingly finer in later stages''.\footnote{Specific details or the code are not given in the paper.} Adaptive support approach is also used in more recent computational methods of \cite{YBT}, \cite{PZ} and \cite{Duarte}. The algorithms by \cite{YBT} and \cite{PZ} use classical convex optimization for finding optimal weights on a fixed discrete set, and, in each step, they enrich the support set with the global maximum of the variance function. Finally, \cite{Duarte} suggest to use semidefinite programming as a discrete-space solver and iteratively replace the support by a specific set of local maximizers of the variance function.}

\section{The algorithm}\label{sec:TheAlgorithm}

In GEX, we apply a finite-space optimum design algorithm on suitably chosen exploration sets $\X_{exp}$. In the course of the computation, the exploration sets are sequentially updated with the aim to approach the support of the true optimal design for the problem at hand. An outline of GEX is exhibited as Algorithm \ref{algo:GEXout} below. The input of the algorithm involves the problem itself in the form of a subroutine that for each permissible $\x \in \X$ computes the vector $\f(\x)$. The values $\eff_{opt}<1$, $\eff_{stop}<1$, $\eff_{grp} \leq 1$ and $N_{loc} \in \{0,1,\ldots\}$ are parameters chosen by the user; their meaning is explained in the sequel.
\begin{algorithm}[]
	\SetAlgoLined
	\DontPrintSemicolon
	\KwIn{$\X$, $\f$, $\eff_{opt}$, $\eff_{grp}$, $\eff_{stop}$, $N_{loc}$}
	\nlset{1}$\X_{exp} \leftarrow$ \texttt{INI} ($\X$) \\
	\nlset{2}$\xi_{new} \leftarrow$ \texttt{OPT} ($\X_{exp}$, $\f$, $\eff_{opt}$, $\eff_{grp}$) \\
	\nlset{3}\Repeat{$\phi(\xi_{old})/\phi(\xi_{new}) > \eff_{stop}$}{
	\textbf{\footnotesize(a)} $\xi_{old} \leftarrow \xi_{new}$ \\
	\textbf{\footnotesize(b)} $\X_{exp} \leftarrow$ \texttt{EXP} ($\X$, $\f$, $\xi_{new}$, $N_{loc}$) \\
	\textbf{\footnotesize(c)} $\xi_{new} \leftarrow$ \texttt{OPT} ($\X_{exp}$, $\f$, $\eff_{opt}$, $\eff_{grp}$)
	}
	\nlset{4}\KwRet{$\xi_{new}$}
	\caption{\textbf{GEX (outline)}}\label{algo:GEXout}
\end{algorithm}
 
In Steps 1 and 2 we select an initial finite exploration set $\X_{exp} \subseteq \X$, and use a finite-space method to construct an $\eff_{opt}$-efficient design for the model determined by $\f$ with the design space restricted to $\X_{exp}$. In Step 3 we alternate the construction of a finite exploration set $\X_{exp}$ based on the current design and the computation of a new $\eff_{opt}$-efficient design on $\X_{exp}$. The algorithm stops once the last optimization on $\X_{exp}$ does not lead to a significant improvement. Note that this form of the stopping rule implies that the number of loops in Step 3 is bounded.   
\bigskip

Although the general scheme of GEX is simple, a judicious specification of the basic steps can make a crucial difference in the overall performance. In the following subsections we detail our choice of the procedures \texttt{INI}, \texttt{OPT} and \texttt{EXP}.

\subsection{\texttt{INI}}\label{ssINI}
 
In Step 1 of Algorithm \ref{algo:GEXout}, we construct an initial exploration set $\X_{exp} \subseteq \X$. In our specification, $\X_{exp}$ is constructed as a union of $\X_{grid}$ and $\X_{rnd}$. The set $\X_{grid}$ is a grid formed by the combinations of the extreme levels of all factors and the median levels of all non-binary factors. The set $\X_{rnd}$ is a random selection of points in $\X$. The size of $\X_{grid}$ is at most $3^k$, which is reasonably small for as many as $k \lesssim 13$ factors, and, for many models used in practice, $\X_{grid}$ constructed in the proposed way contains highly informative points. For our numerical study we chose the size of $\X_{rnd}$ to be $1000$; a sensitivity analysis suggests that the performance of the algorithm is similar for a wide range of sizes of $\X_{rnd}$.    

\subsection{\texttt{OPT}}\label{ssOPT}

In Steps 2 and 3(c) of Algorithm \ref{algo:GEXout}, we apply a discrete-space algorithm to compute an $\eff_{opt}$-efficient design on $\X_{exp}$ with a support pruned by a grouping procedure. In more detail, the assignment $\xi_{new} \leftarrow$ \texttt{OPT} ($\X_{exp}$, $\f$, $\eff_{opt}$, $\eff_{grp}$) consists of the steps shown below.
\RestyleAlgo{plainruled}
\begin{algorithm}[]
	\SetAlgoLined
	\DontPrintSemicolon
	\Begin({\texttt{OPT}($\X_{exp}$, $\f$, $\eff_{opt}$, $\eff_{grp}$) }){
		 $\xi_{new} \leftarrow$ \texttt{REX} ($\X_{exp}$, $\f$, $\eff_{opt}$) \\
		 $\xi_{new} \leftarrow$ \texttt{GRP} ($\xi_{new}$, $\f$, $\eff_{grp}$)
	}
\end{algorithm}
\RestyleAlgo{algoruled}

As the underlying finite-space optimization procedure we chose the REX algorithm from \cite{REX}. It has several advantages compared to other methods: REX is not only fast and applicable to relatively large problems, but, crucially, the resulting designs do not have the tendency to contain many design points with small ``residual'' weights. The parameter $\eff_{opt}$ is the lower bound on the efficiency to stop the computation. In the numerical studies we chose $\mathrm{eff}_{opt} = 1 - 10^{-6}$.

Note that in Step 2 of Algorithm \ref{algo:GEXout}, the initial design for REX is computed via the modified Kumar-Yildirim method as described in \cite{PIN}, which is rapid and always provides a design with a non-singular information matrix. If this more advanced initiation method is replaced by a uniformly random selection of $m$ distinct points, the efficiency of the resulting design (as well as the speed of computation) is not much affected. However, the Kumar-Yildirim method guarantees non-singularity of the information matrix of the initial design which is an important aspect of the algorithm.\footnote{In many models used in practice the probability that a random $m$-point design is singular is significantly greater than zero.} In contrast to Step 2, in Step 3(c) the REX algorithm is initialized via $\xi_{new}$.

Each computation of REX is followed by a specific form of grouping of nearby support points. For continuous spaces, the clustering of nearby points was proposed already in \cite{Fedorov}, and there are many straightforward variants. In our case, the factors are discrete, but the levels can be very close to each other, which means that grouping of nearby support points generally improves the performance. We identified that for all studied models it is enough to use a nearest-distance approach to decrease the support size of the constructed design on a discrete space. 

More precisely, let $\xi_0$ be the design resulting from the the original discrete-space procedure. Let $\x_1,\ldots,\x_s$ be the support points of $\xi_0$.\footnote{Note that using REX as the engine provides the designs with relatively small support sizes, almost always smaller than $m^2$.} The procedure \texttt{GRP} determines the pair $(\x_k,\x_l)$, $k<l$, of two nearest support points and assigns the pooled weight $\xi_0(\x_k)+\xi_0(\x_l)$ to the point $\x_k$ (if $\xi_0(\x_k) \geq \xi_0(\x_l)$) or to the point $\x_l$ (if $\xi_0(\x_k) < \xi_0(\x_l)$), which results in a design $\xi_1$. If $\xi_1$ is at least $\mathrm{eff}_{grp}$-efficient relative to the design originally returned by REX, the pooling operation is accepted and the process is repeated. In our computation study, we chose $\mathrm{eff}_{grp}=1-10^{-6}$. Note that removing \texttt{GRP} (e.g., by setting $\mathrm{eff}_{grp}=1$) provides negligibly more efficient designs, but it makes the computation slower. More importantly, not using \texttt{GRP} results in designs that are populated by a large number of points with small weights.

\subsection{\texttt{EXP}}\label{ssEXP}

In the key Step 3(b) we construct a new exploration set $\X_{exp}$ based on a set of local maxima of the variance function, and a long-range variation of the support of the current design. In detail, $\X_{exp} \leftarrow$ \texttt{EXP} ($\X$, $\f$, $\xi_{new}$, $N_{loc}$), consists of the part summarized below. 
\RestyleAlgo{plainruled}
\begin{algorithm}[]
	\SetAlgoLined
	\DontPrintSemicolon
	\Begin({\texttt{EXP}($\X$, $\f$, $\xi_{new}$, $N_{loc}$)}){
	$\X_{loc} \leftarrow$ \texttt{LOC} ($\X$, $\f$, $\xi_{new}$, $\mathrm{N}_{loc}$) \\
	$\X_{star} \leftarrow$ \texttt{STAR} ($\X$, $\xi_{new}$)  \\
	$\X_{exp} \leftarrow \X_{loc} \cup \X_{star}$
	}
\end{algorithm}
\RestyleAlgo{algoruled}

The procedure \texttt{LOC} returns the result of a randomly initialized local-search greedy maximization of the current variance function $d_{\xi_{new}}(\cdot)$ over $\X$. The number of greedy optimizations starting from random points of $\X$ is given by the parameter $\mathrm{N}_{loc}$; for our numerical study we chose $\mathrm{N}_{loc} = 50$. Decreasing $N_{loc}$ slightly worsens the resulting designs but also makes the computation somewhat faster. Removing the $\X_{loc}$ part of $\X_{exp}$ altogether often makes the computation much faster, but in some cases it leads to systematically and significantly suboptimal designs. 

As in all local search methods, the efficiency depends on the system of neighbourhoods for each feasible solution $\x$. 

We use ``star set'' neighborhoods $\mathfrak{S}(\x)$ which consist of all points lying in a ``star'' centred at the considered point $\x$: $\mathfrak{S}(\x)$ is the set of points in $\X$ that differ from $\x$ in at most one coordinate. Formally,
\begin{equation*}
 \mathfrak{S}(\x)=\left\{\tilde{\x} \in \X \, \vert \, \exists i \in \{1,\ldots,k\}: \tilde{x}_i \in \X_i, \tilde{x}_j = x_j \, \forall j \neq i \right\}.
\end{equation*}
We search the neighbourhood of the current point $\x$, move to the point in $\mathfrak{S}(\x)$ with the highest value of the variance function and repeat while there is an increase in the variance function by moving to the new point. The neighbourhoods $\mathfrak{S}(\x)$ seem to be suitable for optimal design problems on multidimensional grids as they allow to explore the grids quite thoroughly, yet they do not grow exponentially in size with respect to the dimension $k$ of the design space. Moreover, the neighbourhoods $\mathfrak{S}(\x)$ have the added advantage that they are fast to computationally enumerate.
\bigskip

\begin{figure}
	\centering
	\includegraphics[width=0.5\textwidth]{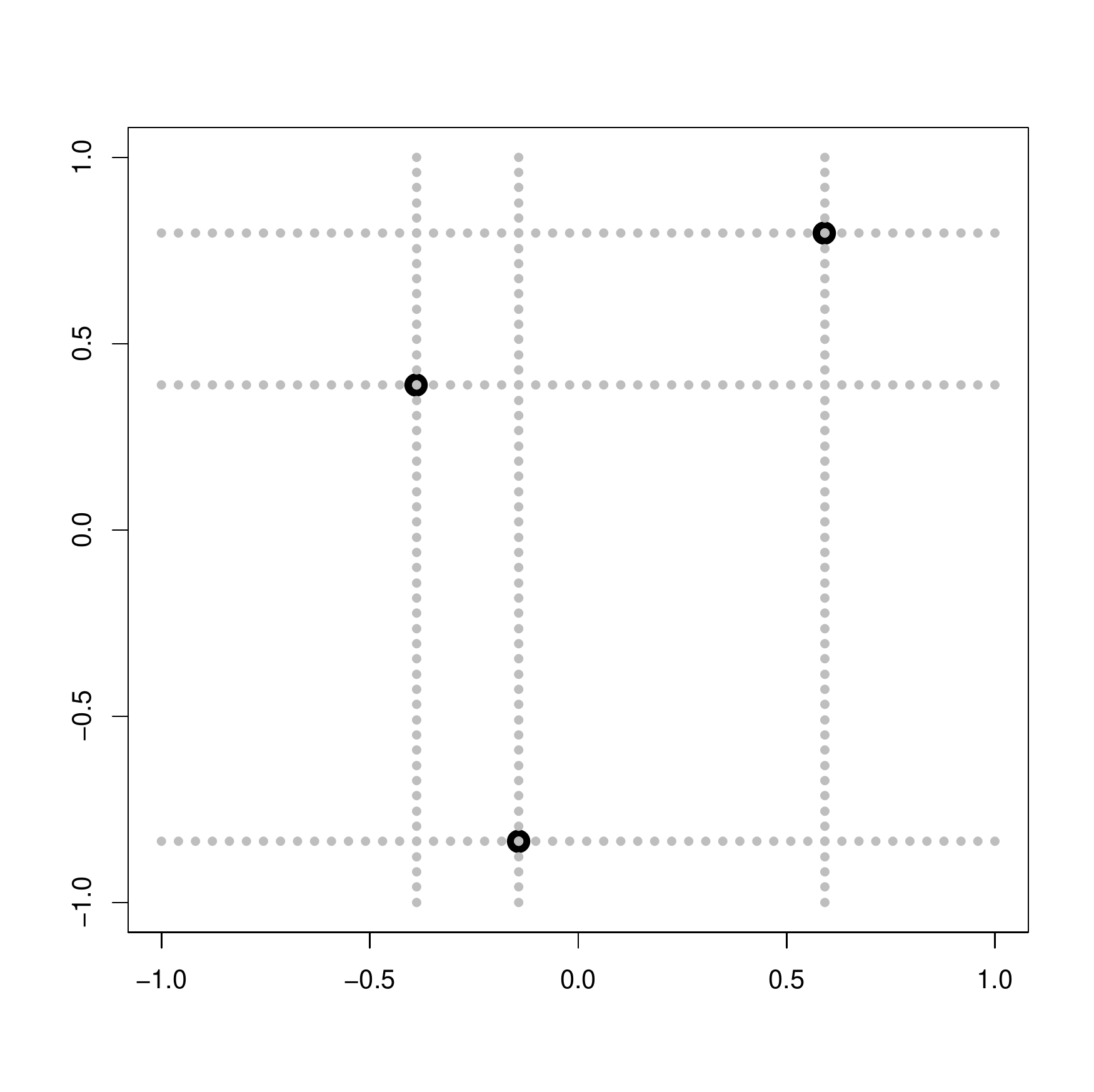}
	\caption{$\X_{star}$ for a design with $3$ support points, where $\X$ is a discretized square $[-1,1]^2$. The support points are denoted by black circles, and the grey dots form the star sets.}
	\label{fStar}
\end{figure}

The crucial procedure for the construction of the exploration sets is \texttt{STAR}, which is specifically chosen to fit the grid-like structure of the design space studied in the paper. The best solution requires a balanced compromise between exploration and exploitation. Our experience leads us to choose star sets \emph{again} (hence, the name Galaxy EXploration). For a given $\xi_{new}$, the set $\X_{star}$ is the union of all star-set neighbourhoods centred at the support points of $\xi_{new}$. Formally
\begin{equation*}
 \X_{star} = \bigcup_{\x \in \mathrm{supp}(\xi_{new})} \mathfrak{S}(\x);
\end{equation*}
such a star set is illustrated in Figure \ref{fStar}. 
Note that the REX algorithm produces efficient designs with ``sparse'' supports, which means that the size of $\X_{star}$ does not exceedingly grow during the computation of GEX.

\subsection{Summary}\label{sec:Summary}

Algorithm \ref{algo:GEXdetail} shows the final formulation of GEX, with all the steps described in detail.
\begin{algorithm}[]
	\SetAlgoLined
	\DontPrintSemicolon
	\KwIn{$\X$, $\f$, $\eff_{opt}$, $\eff_{grp}$, $\eff_{stop}$, $\mathrm{N}_{loc}$}
	\nlset{1}$\X_{exp} \leftarrow$ \texttt{INI} ($\X$) \\
	\nlset{2}$\xi_{new} \leftarrow$ \texttt{OPT} ($\X_{exp}$, $\f$, $\eff_{opt}$) \\
	\nlset{3}\Repeat{$\phi(\xi_{old})/\phi(\xi_{new}) > \eff_{stop}$}{
		\nlset{3a} $\xi_{old} \leftarrow \xi_{new}$ \\
		\nlset{3b} \Begin({\texttt{EXP}}){
			$\X_{loc} \leftarrow$ \texttt{LOC} ($\X$, $\f$, $\xi_{new}$, $\mathrm{N}_{loc}$) \\
			$\X_{star} \leftarrow$ \texttt{STAR} ($\X$, $\xi_{new}$)  \\
			$\X_{exp} \leftarrow \X_{loc} \cup \X_{star}$
		}
		\nlset{3c}\Begin({\texttt{OPT}}){
			$\xi_{new} \leftarrow$ \texttt{REX} ($\X_{exp}$, $\f$, $\eff_{opt}$) \\
			$\xi_{new} \leftarrow$ \texttt{GRP} ($\xi_{new}$, $\f$, $\eff_{grp}$)
		}
	}
	\nlset{4}\KwRet{$\xi_{new}$}
	\caption{\textbf{GEX (detailed)}}\label{algo:GEXdetail}
\end{algorithm}

Similarly to \cite{PZ} and \cite{Duarte}, our method applies a discrete-space solver to a subset of $\X$, and adaptively modifies this subset. However, unlike \cite{PZ} and many other methods, GEX does not attempt to calculate the global maximum of the variance function, but utilizes a set of local maximizers. The general idea of using a set of local maximizers is similar to \cite{Duarte}, but in GEX we accomplish it in a significantly different way. Importantly, we include the star sets around the support points of the current $\xi$, which means that the exploration set is much bigger than just a set of local maximizers of the variance function, which entails that GEX is able to better explore the design space. Note also that the $\X_{star}$ part of the exploration set is similar to the set explored by the coordinate-exchange method for exact designs (see \cite{MN95}), where in each iteration, one coordinate of one design point is updated via line search. However, we compute approximate, not exact designs, and, crucially, do not consider the coordinates of the points one at a time; rather, all single-coordinate changes of all points are considered simultaneously as a batch (with a set of local maximizers of the variance function), upon which an efficient internal discrete-space solver is applied.
\bigskip

In summary, GEX is practically and numerically simple to apply in the sense that it requires no theoretical analysis of the model at hand, and no special numerical routines, solvers, or libraries, except for the standard procedures of linear algebra. The algorithm only has a few tuning parameters, which are directly interpretable and can be set to fixed default values that perform well for a wide class of problems (as we demonstrate in the next section). Moreover, GEX automatically determines the support size, i.e., no search for the proper size of the support is needed.

On the other hand, GEX cannot be applied if we do not have a discrete-space solver for the required criterion. Nevertheless, for the most common criteria, such that $D$-, $A$-, $I$- and $c$-optimality, there is a large selection of suitable discrete-space methods. Finally, GEX does not necessarily converge; note, however, that no optimal design method can be claimed to always converge, unless it inspects \emph{all} design points at least once (which is simply impossible for larger problems) or makes use of specific analytic properties of the model. 

\subsection{Notes}\label{ssNotes}

In some cases, the information matrix of the $D$-optimal design is badly conditioned. However, we can use the well-known fact that a regular re-parametrization, i.e., replacing the vectors $\f(\x)$ with $\tilde{\f}(\x)=\mathbf{R}\f(\x)$, where $\mathbf{R}$ is non-singular, does not alter the $D$-optimal design. If $\mathbf{R}$ is chosen to be close to the inverse of the square root of the optimal information matrix, such a re-parametrization may greatly improve the condition number of the matrices used by GEX. We applied this approach to increase the numerical stability of the computation of the problem number 10 described in Section \ref{sec:Study}. We chose $\mathbf{R}=\mathbf{M}^{-1/2}$, where $\mathbf{M}$ is the information matrix of the design computed at the initial $\X_{exp}$ by the modified Kumar-Yildirim method.  

During the run of the algorithm, the points that cannot support optimal designs can be removed using the results of \cite{del} to decrease the size of the problem and potentially speed up the computations. However, in our experience, the removal of redundant points does not provide a significant increase in the speed of GEX. This is likely due to the nature of the finite-space engine REX, which tends to work with designs of small supports, and as such does not benefit much from the reduction of the unpromising parts of the design space.

\section{A numerical study}\label{sec:Study}

For a numerical study, we selected $5$ prominently published recent papers that focus on presenting a method for computing $D$-optimal designs on continuous or mixed design spaces, namely \cite{PZ}, \cite{Duarte}, \cite{Lukemire}, \cite{Xu}, \cite{Garcia-Rodenas}. From each paper, we chose two test problems, which are summarized in Table \ref{tab:problems}. The problems include a linear regression model, two nonlinear models with homoscedastic normal errors, a Poisson, probit, and several logistic regression models. The number $k$ of factors varies from $2$ to $10$, and the parametric dimension $m$ from $4$ to $16$. We used these test problems to illustrate the behaviour of GEX and numerically compare its performance to the competing methods.

\begin{figure*}
    \includegraphics[width=0.95\textwidth]{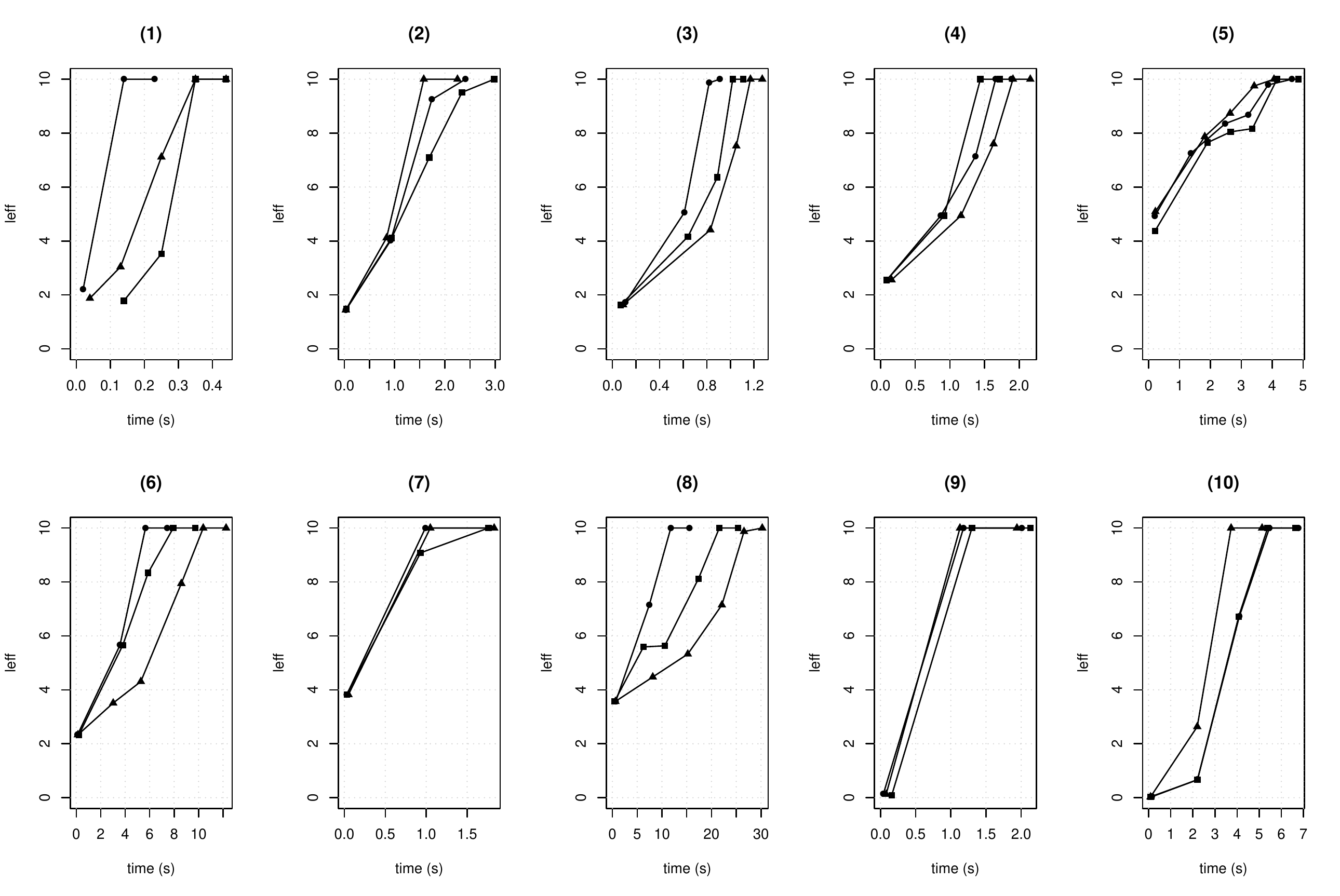}
\caption{The time profile of the efficiencies of designs $\xi_{new}$ computed by REX at the Steps 2 and 3(c) of GEX. For each problem (see the numbers above panels), we executed $3$ independent runs of GEX; each computation is denoted by a separate piecewise-linear curve. The efficiencies are expressed in a logarithmic scale as follows: The efficiencies $0.9$, $0.99$, $0.999$, ..., $1-10^{-9}$, $>1-10^{-9}$ correspond to the values $1$, $2$, ..., $9$, $10$ at the vertical axis.}\label{fig:GEXperf}
\end{figure*}

Figure \ref{fig:GEXperf} shows the typical behaviour of GEX on the test problems from Table \ref{tab:problems}. We see that the designs $\xi_{new}$ generated in the main loop of GEX monotonically improve, and ultimately converge to the optimum.\footnote{By optimal design we mean the best design that we are aware of from our numerical studies or from the literature. In addition, while we cannot prove that GEX always converges to the perfect optimum, we observed the convergence to the same design in all runs.} The variability of the computation time is moderate.\footnote{The coefficient of variation of the computation times is less than $0.3$ in all studied problems.} The usual number of discrete-space computations is 3 to 6 before the stopping rule is satisfied. Note that the efficiency growth generally does not decline with computation time. With increasing difficulty of the problem, the number of computations required to achieve the optimal design does not tend to increase.

We provide \texttt{R} (\cite{R}) codes which are completely free to use and require minimum technical expertise to be applied to any particular optimal design problem of the reader, see

\begin{small}
\url{http://www.iam.fmph.uniba.sk/ospm/Harman/design/}
\end{small}
\bigskip

As an illustration, in Table \ref{tab:design6} we exhibit the design for Problem 6 as computed by GEX. All other designs can be obtained within a few seconds by running the provided \texttt{R} script. Note that from the nature of GEX the resulting designs are ``tidy'', i.e., there is no need to remove the design points with very small weights or round the positions of the support points to a reasonable number of decimal places.  

\begin{table}
\caption{The design produced by GEX for Problem 6. The first column is the index $i$ of the identified support point, the columns $x_{i1}$ to $x_{i5}$ are the values of the factors and $\xi(\x_i)$ is the weight of the corresponding support point $\x_i$.}\label{tab:design6}
\begin{tabular}{rrrrrrr}
\hline\noalign{\smallskip}
$i$ & $x_{i1}$ & $x_{i2}$ & $x_{i3}$ & $x_{i4}$ & $x_{i5}$ & $\xi(\x_i)$\\
\noalign{\smallskip}\hline\noalign{\smallskip}
1  & -2.000  & -2 & -2 & -2.000 & -2 & 0.093630\\
2  & -2.000  &  2 &  2 & -2.000 & -2 & 0.092334\\
3  &  2.000  & -2 & -2 &  2.000 & -2 & 0.062242\\
4  &  2.000  &  2 & -2 &  2.000 & -2 & 0.075893\\
5  &  2.000  &  2 &  2 &  2.000 & -2 & 0.065934\\
6  & -2.000  & -2 &  2 & -2.000 &  2 & 0.076828\\
7  & -2.000  &  2 & -2 &  2.000 &  2 & 0.089598\\
8  & -0.930  &  2 & -2 & -2.000 & -2 & 0.080512\\
9  & -1.739  & -2 & -2 &  2.000 &  2 & 0.060395\\
10 & -2.000  & -2 & -2 & -1.340 &  2 & 0.024540\\
11 & -0.528  &  2 & -2 &  2.000 &  2 & 0.004427\\
12 & -1.636  &  2 &  2 &  2.000 &  2 & 0.054245\\
13 &  2.000  & -2 &  2 &  1.838 & -2 & 0.055518\\
14 &  1.793  & -2 &  2 &  2.000 &  2 & 0.083511\\
15 & -2.000  &  2 &  2 & -1.483 &  2 & 0.034830\\
16 &  1.391  & -2 &  2 & -2.000 & -2 & 0.045563\\
\noalign{\smallskip}\hline
\end{tabular}
\end{table}

In optimal experimental design, numerical comparison of competing computational methods is a challenging task. The reason is that there are no generally adopted benchmark suites and no guidelines for reporting the results of comparison. Often, the computer code is not available or is fine-tuned to the few problems studied in the paper. Moreover, the quality of the resulting designs depends on the computation time in a method-specific way and can be strongly influenced by the choice of the hardware, programming language and implementation details; a seemingly minor change can lead to a severalfold speed-up or slow-down. Despite the methodological difficulties, at least a brief comparison is necessary, because there is a plethora of heuristic or theoretical ideas applicable to computing optimal designs and failing to realize that a proposed method is much worse than existing alternatives can be harmful to the readers.
\bigskip

In Table \ref{tab:results}, we provide the computation times and the criterion values of the optimal or nearly-optimal designs, as reported in the corresponding papers\footnote{In selected cases, we used our own hardware to re-compute the designs; see the following text.} and as provided by an \texttt{R} implementation of GEX. We used the Microsoft \texttt{R} Open 3.5.3 and a 64-bit Windows 10 system with an Intel Core i7-9750H processor at 2.60 GHz. 

The authors of the competing methods used either Matlab or C++, which almost always allows for a faster implementation of a given algorithm than \texttt{R}. Except for \cite{PZ}, the authors used standard modern hardware. 

The results for models 2 and 3 were obtained by running $11$ times the algorithm of \cite{PZ} using our hardware and the Matlab code kindly provided by the authors. For these two models, the authors also give analytically calculated optimal designs; thus, it can be shown that the designs obtained by GEX for these models are at least 99.999\%-efficient relative to the provably optimal designs. In models 1 and 4, the CPU times and the optimal designs were taken from Table 10 in \cite{Duarte}. The results of models 5 and 8 from \cite{Lukemire} were obtained by running the compiled instance of the program written in C++ provided by the authors, with the number of iterations set to $2000$ and the required number of support points pre-determined by GEX. We ran each instance $11$ times and provided the computation times and values of the criterion for the resulting designs. The results of \cite{Xu} for model 9 can be found in Section IV and the results for model 10 are taken from Section V, part A. The computation times are based on the personal communication with the lead author of \cite{Xu}. Finally, the results for models 6 and 7 are available in Table 11 of \cite{Garcia-Rodenas}; note that the exhibited results encompass the time and efficiency performance of $5$ different algorithms.
\bigskip

From Table \ref{tab:results}, it is clear that GEX produces designs with higher $D$-efficiency than competing methods, in a shorter computation time, sometimes by several orders of magnitude. The only close competitor is the algorithm from \cite{PZ}. In fact, the designs for problems 2 and 3 obtained by this algorithm are slightly better than the designs for the same problems obtained by GEX, although the difference is not within $6$ significant digits of the criterion value displayed in Table \ref{tab:results}. The practical use of the algorithm from \cite{PZ} is, however, limited to the problems with small numbers of factors (if the number of factors is more than $2$, the computation becomes exceedingly slow). In addition, in contrast to GEX, the application of this algorithm may require theoretical analysis specific to the problem. 

However, we stress that all compared algorithms have their own advantages, for instance they may be more appropriate for computing designs efficient with respect to special optimality criteria. 

\begin{table*}
	\caption{The description of the benchmark models, the nominal parameter $\theta_0$ used and the design space $\X$. The level of discretization of the continuous intervals is denoted in the subscript. For instance $[-1,1]^2_{0.001}$ denotes two factors ranging from -1 to 1, each of them discretized with step $0.001$. The columns $k$ and $m$ summarize the number of factors and the number of parameters, respectively.}\label{tab:problems}
	\begin{tabular}{llllll}
		\hline\noalign{\smallskip}
		\# & model & $\theta_0$ & design space $\X$ & $k$ & $m$\\
		\noalign{\smallskip}\hline\noalign{\smallskip}
		1 & $y(\x) \sim N(\eta, \sigma^2)$, $\eta=\frac{1}{1+e^{\h^T(\x)\theta}}$,  & $(-2, 0.5, 0.5, 0.1)^T$ & $[0,5]_{0.001}\times [0,1]_{0.001}$ &2 & 4 \\
		& $\h^T(\x)=(1,x_1,x_2,x_1x_2)$ &&&&\\[2mm]
		2 & $y(\x) \sim N(\eta, \sigma^2)$,  & $(1, 1, 2, 0.7, 0.2)^T$ & $[0,2]_{0.001}\times [0,10]_{0.001}$ & 2 & 5\\
	    &$\eta=\theta_1 + \theta_2 e^{-\theta_3 x_1}+ \frac{\theta_4}{\theta_4 - \theta_5} (e^{-\theta_5 x_2}-e^{-\theta_4 x_2})$&&&&\\[2mm]
		3 & $y(\x) \sim N(\eta, \sigma^2)$, $\eta=\h^T(\x)\theta$,
		 & does not affect the design & $[-1,1]_{0.001}^2$ & 2 & 7 \\
	    &$\h^T(\x) = (1, x_1, x_2, x_1^2, x_2^2, x_1^3, x_2^3)$&&&&\\[2mm]
		4 & $y(\x) \sim Pois(\eta)$, $\eta=e^{\h^T(\x)\theta}$,  &$(0.5, -0.2, 0.5, -0.2, -0.1, $&$[-1,1]_{0.001}^3$& 3 & 10\\
		&$\h^T(\x)=(1,\x^T,x_1^2, x_2^2, x_3^2,x_1x_2,x_1x_3,x_2x_3)$& $0.2, -0.1, 0.2, -0.1, 0.2)^T$ &&&\\[2mm]
		5 & $y(\x) \sim Bin(1, \eta)$, $\eta=\frac{e^{\h^T(\x)\theta}}{1+e^{\h^T(\x)\theta}}$,  & $(-1, 2, 0.5, -1, -0.25, 0.13)^T$ & $\{-1,1\}^4\times [5,35]_{0.001}$& 5 & 6 \\
		&$\h^T(\x)=(1,\x^T)$&&&&\\[2mm]
		6 & $y(\x) \sim Bin(1,\eta)$, $\eta=\phi(\h^T(\x)\theta)$, & $(0.5, 0.7, 0.18, -0.2, -0.58, 0.51)^T$ & $[-2,2]_{0.001}^5$ & 5 & 6 \\
		& $\h^T(\x)=(1,\x^T)$&&&&\\[2mm]
		7 & $y(\x) \sim Bin(1,\eta)$, $\eta=\frac{e^{\h^T(\x)\theta}}{1+e^{\h^T(\x)\theta}}$,  & $(0.5, 0.7, 0.18, -0.2, -0.58, 0.51)^T$ & $[-2,2]_{0.001}^5$ & 5 & 6\\
		&$\h^T(\x)=(1,\x^T)$&&&&\\[2mm]
		8 & $y(\x) \sim Bin(1,\eta)$, $\eta=\frac{e^{\h^T(\x)\theta}}{1+e^{\h^T(\x)\theta}}$,  & $(-0.4926, -0.628, -0.3283, 0.4378,$ & $[-3,3]_{0.01}^7$ & 7 & 8 \\
		&$\h^T(\x) = (1, \x^T)$& $0.5283, -0.612, -0.6837, -0.2061)^T$&&&\\[2mm]
		9 &  $y(\x) \sim Bin(1,\eta)$, $\eta=\frac{e^{\h^T(\x)\theta}}{1+e^{\h^T(\x)\theta}}$,  & $(3, 0.5, 0.75, 1.25, 0.8, 0.5, 0.8, $ & $\{-1,1\}^4\times[50,90]_{0.01}\times [30,55]_{0.01}$ & 10 & 11\\
		&$\h^T(\x)=(1,\x^T)$&$-0.4,-1.0, 2.65, 0.65)^T$& $\times[0,10]_{0.01}\times[18,48]_{0.01}$ &&\\
		&&& $\times[0.125,0.425]_{0.001}\times[5,15]_{0.01}$ &&\\[2mm]
		10 & $y(\x) \sim Bin(1,\eta)$, $\eta=\frac{e^{\h^T(\x)\theta}}{1+e^{\h^T(\x)\theta}}$,   & $(3, 0.5, 0.75, 1.25, 0.8, 0.5, 0.8, $ & $\{-1,1\}^4\times[50,90]_{0.01}\times[30,55]_{0.01}$ & 10 & 16\\ 
		&$\h^T(\x) = (1, \x^T, x_1x_9, x_2x_5, x_3x_4, x_6x_7, x_8x_{10})$ & $-0.4, -1.0, 2.65, 0.65,0.01, $ & $\times[0,10]_{0.01}\times[18,48]_{0.01}$ &&\\
		&&$ -0.02, 0.03, -0.04, 0.05)^T$ & $\times[0.125,0.425]_{0.01}\times[5,15]_{0.01}$ &&\\
		\noalign{\smallskip}\hline
	\end{tabular}
\end{table*}

\begin{table*}
	\caption{The performance of GEX and the competing methods on models from Table \ref{tab:models}. The first column represents the model number, $t_{GEX}$ is the minimum and maximum computation time out of $11$ runs of GEX, and $\Phi^*_{GEX}$ is the obtained criterion value (for each model, we obtained the same criterion value in each run). The columns $t_{COM}$ and $\Phi^*_{COM}$ represent the computation times and the obtained criterion values, respectively, for the competing method (column ``source''). A more detailed description of these values is given in the text.}\label{tab:results}
\centering
	\begin{tabular}{llllll}
		\hline\noalign{\smallskip}
		\# & $t_{GEX}$ (sec) & $\Phi^*_{GEX}$ & source & $t_{COM}$ (sec) & $\Phi^*_{COM}$\\
		\noalign{\smallskip}\hline\noalign{\smallskip}
		2 & 2.20 - 3.67 & 0.117578 & \cite{PZ} & 3.68 - 3.92 & 0.117578 - 0.117578\\
		3 & 0.96 - 1.51 & 0.221567 & \cite{PZ} & 3.05 - 3.33 & 0.221567 - 0.221567\\[2mm]
		1 & 0.22 - 0.45 & 0.0338935 & \cite{Duarte} & 12.00 & 0.0338904 \\
		4 & 1.74 - 2.25 & 0.870542 & \cite{Duarte} & 72.86 & 0.853086 \\[2mm]
		5 & 2.84 - 5.19 & 0.351996 & \cite{Lukemire} & 61.05 - 63.76 & 0.348210 - 0.350217 \\
		8 & 2.05 - 3.06& 0.0381948 & \cite{Lukemire} & 257.89 - 321.38 & 0.0378399 - 0.0381872\\[2mm]
		9 & 15.04 - 28.63& 1.07287 & \cite{Xu} & $\leq 600$ & 1.06962 \\
		10 & 4.96 - 6.89& 0.0115145 & \cite{Xu} & $\leq 60$ & 0.0114329 \\[2mm]
		6 & 7.53 - 12.95& 1.26609 & \cite{Garcia-Rodenas} & 34.36 - 225.24 & 1.05263 - 1.23457 \\
		7 & 1.86 - 2.88& 0.539359 & \cite{Garcia-Rodenas} & 9.82 - 83.23 & 0.462963 - 0.526315 \\
		\noalign{\smallskip}\hline
	\end{tabular}
\end{table*}

\section{Final comments}\label{sec:conclusions}

We proposed a conceptually simple approach and its concrete and efficient specification which we call the Galaxy exploration method (GEX). The algorithm can be used for computing optimal or nearly-optimal approximate experimental designs on large grids, and, by means of a dense discretization, on multidimensional cuboid spaces. Note that utilizing a suitable transformation, GEX can also be applied to computing efficient designs on different design spaces. For instance, the surface or the interior of a ball can be transformed to a cuboid by means of polar coordinates.

The performance of GEX can be undoubtedly improved even further, for instance by a modification of the exploration sets, clever adaptive changes of the levels of each factor, or by parallelization of the computation.  

\begin{acknowledgements}
The work was supported by Grant No. 1/0341/19 from the Slovak Scientific Grant Agency (VEGA). We are also grateful to Dr. Luc Pronzato and prof. Anatoly Zhigljavsky for the Matlab codes implementing their algorithm from the paper \cite{PZ}. We also thank Dr. Xu Weinan for the information on the computation of the designs from the publication \cite{Xu}.
  
\end{acknowledgements}



\end{document}